\documentclass[amssymb,useAMS,prd,aps,twocolumn,superscriptaddress]{revtex4}
\usepackage{pslatex}
\usepackage{graphicx}
\usepackage{psfrag}
\usepackage{epsfig}

\def\refe@jnl#1{{#1}}
\def\aj{\refe@jnl{Astron.~J.}}
\def\araa{\refe@jnl{Annu.~Rev.~Astron.~Astrophys.}}
\def\apj{\refe@jnl{Astrophys.~J.}}
\def\apjl{\refe@jnl{Astrophys.~J.~Lett.}}
\def\aap{\refe@jnl{Astron.~Astrophys.}}
\def\mnras{\refe@jnl{Mon.~Not.~R.~Astron.~Soc.}}
\def\prd{\refe@jnl{Phys.~Rev.~D}}
\def\fcp{\refe@jnl{Fund.~Cos.~Phys.}}
\def\physrep{\refe@jnl{Phys.~Rep.}}
\def\physlett{\refe@jnl{Phys.~Lett.}}

\def\invisible#1{  }

\def\lsim{\mathrel{\lower4pt\hbox{$\sim$}} 
\hskip-9.5pt\raise1.6pt\hbox{$<$}\;} 
 
\def\gsim{\mathrel{\lower4pt\hbox{$\sim$}} 
\hskip-9.5pt\raise1.6pt\hbox{$>$}\;}

\begin{document}
\title{WIMP dark matter, Higgs exchange and DAMA}

\author{Sarah Andreas}
\affiliation{Service de Physique Th\'eorique\\
 Universit\'e Libre de Bruxelles\\ 
Boulevard du Triomphe, CP225, 1050 Brussels, Belgium\footnote{Sarah.Andreas@rwth-aachen.de; thambye@ulb.ac.be; mtytgat@ulb.ac.be}}
\affiliation{Institut f\"ur Theoretische Physik E\\RWTH Aachen University, D-52056 Aachen, Germany}
\author{Thomas Hambye} 
\author{Michel H.G. Tytgat}
\affiliation{Service de Physique Th\'eorique\\
 Universit\'e Libre de Bruxelles\\ 
Boulevard du Triomphe, CP225, 1050 Brussels, Belgium\footnote{Sarah.Andreas@rwth-aachen.de; thambye@ulb.ac.be; mtytgat@ulb.ac.be}}

%\date{\today}

\begin{abstract}

In the WIMP scenario, there is a one-to-one relation between the dark matter (DM) relic density and spin independent direct detection rate
if both the annihilation of DM
and 
its elastic scattering on nuclei go dominantly through  Higgs exchange.  In particular, for  DM masses much smaller than the Higgs boson mass,
the ratio of the relevant cross sections depends only on the DM mass.
Assuming  DM mass and direct detection rate within the ranges allowed by the recent DAMA collaboration results --taking account of the channelling effect
 on energy threshold and the  null results of the other direct detection experiments-- gives a definite range for the relic density. For scalar DM models,
like the Higgs portal models or the inert doublet model,
the relic density range turns out to be in agreement with WMAP.
This scenario implies that the Higgs boson has a large branching ratio to pairs of  DM particles, a prediction which might challenge its search at the LHC.

\end{abstract}

 \maketitle
%%%%%%%%%%%%%%%%%%%%%%%%%%%%%%%%%%%%%%%%%%%%%%%%%%%%%%%%%%%%%%%%%%%%%%
%\section{Introduction}
%%%%%%%%%%%%%%%%%%%%%%%%%%%%%%%%%%%%%%%%%%%%%%%%%%%%%%%%%%%%%%%

The DAMA collaboration has recently provided evidence  for an annual modulation of the rate of nuclear recoils in their detector \cite{DAMA}, confirming at 
a firmer level their previous results \cite{DAMA03}.
Taking into account the null results of the other direct Dark Matter (DM) detection experiments \cite{CDMSXenon} and the recently discovered channelling effect on the threshold energy in DAMA, 
points towards 
a nuclear recoil  due to  dark matter-nucleon elastic scatterings, with a spin independent (SI) cross section in the 
range (see~\cite{Petriello:2008jj}) 
\begin{equation}
\label{SI}
3 \times 10^{-41}\, \hbox{cm}^2 \lesssim \sigma_p^{SI} \lesssim 5 \times 10^{-39} \, \hbox{cm}^2
\end{equation}
and dark matter mass in the range
\begin{equation}
\label{mass}
3\,\hbox{GeV}\lesssim m_{DM} \lesssim 8\, \hbox{GeV}.
\end{equation}
These results have been already the object of various studies in specific models \cite{Foot:2008nw,Feng:2008dz,Bottino:2008mf,Avignone:2008cw}. 
In this short letter, working in the WIMP framework which assumes a DM relic density determined by the thermal freeze-out of DM annihilation, we emphasize the importance of Higgs exchange diagrams for DM mass in the range of (\ref{mass}).

If the DM candidate is light, there is a limited number of possible (2-body) annihilation channels to SM particles, {\em i.e.}~to  $u,d,s,c,b$ quark pairs, to
 lepton pairs and to photon pairs. Depending on the model this can be done at tree or loop level in various ways. Annihilations through the SM Z boson may be excluded right away for it would imply that dark matter contributes to the Z invisible width \cite{Yao:2006px}. The next simplest possibility at tree level is annihilation of dark matter into fermions
through the Brout-Englert-Higgs (Higgs for short) in the s-channel, as in the diagram of Fig.1.a. 
In this case, only $b \bar{b}$, $c \bar{c}$ and $\tau \bar{\tau}$ annihilations are relevant, since all other SM fermions have small Yukawa couplings. 
From the very same coupling between DM and the Higgs, SI elastic scattering  is induced through a Higgs in the t-channel, Fig.1.b. 
For the Inert Doublet Model (IDM), a two Higgs model extension of the Standard Model  with 
 a scalar dark matter candidate 
\cite{Deshpande:1977rw,Ma:2006km,Barbieri:2006dq,Majumdar:2006nt,LopezHonorez:2006gr,Gustafsson:2007pc,Hambye:2007vf}, and {\em a fortiori}
 for a singlet scalar DM candidate
\cite{McDonald1,McDonald:2001vt,Burgess:2000yq,Patt:2006fw,Meissner:2006zh,Barger:2007im,Espinosa:2007qk,SungCheon:2007nw},
the  channels of Fig.1 are the only possible non-negligible ones in the range of (\ref{mass}).
In more sophisticated models, such as with the neutralino DM candidate of the MSSM, these channels coexist {\em a priori} with many other channels, 
in particular with other intermediate Higgs scalar particle or squarks channels~\cite{Bottino:2008mf,Bottino:2003iu}.
\begin{center}
\begin{figure}
\vspace{2mm}
\epsfig{file=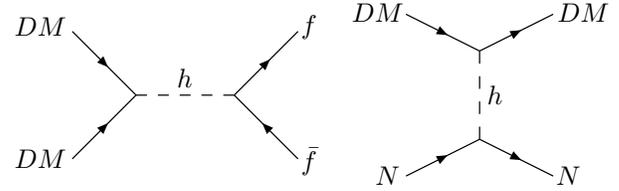,width=8.45cm}
\vspace{-2mm}
\caption{Higgs exchange diagrams for the DM annihilation (a) and scattering with a nucleon (b).}
\vspace{-4mm}
\end{figure}
\end{center}
Both processes in Fig.1 are tightly related. The Higgs boson mass dependence is the same, $m_h^{-4}$, because in both cases the momentum 
of the Higgs is negligible compared with $m_h$. Furthermore the dependence in the unknown DM-DM-$h$ coupling is the same. 
This means that, up to uncertainties in the Yukawa couplings of the SM fermions and in the Higgs to nucleon coupling, the only left parameter is the mass of the dark matter. This dependence is however limited if we take the DM mass within the range (\ref{mass}). 
In other words, if Higgs mediated processes are dominant, the ratio of the cross sections corresponding to the processus of Fig.1 is essentially fixed in any model and there is a one-to-one relation between annihilation and direct detection. This has been pointed out for the case of scalar singlet DM 
in Ref.~\cite{Burgess:2000yq}.
The ratio may however be different in different models. To see this and to inquire whether any model may be in agreement with observations, we consider two simple scenarios, respectively with a scalar and a fermionic dark matter candidate.

For a scalar dark matter candidate,  the simplest possibility is to introduce one real scalar singlet $S$, odd under a $Z_2$ symmetry
\cite{McDonald1,McDonald:2001vt,Burgess:2000yq,Patt:2006fw,Meissner:2006zh,Barger:2007im,Espinosa:2007qk,SungCheon:2007nw},
 like in the so-called Higgs portal framework
\cite{Patt:2006fw,Meissner:2006zh,Barger:2007im,Espinosa:2007qk}.
In full generality  the four following renormalizable terms may be added to the SM lagrangian:
\begin{equation}
\label{lag}
{\cal L} \owns \frac{1}{2}\partial^\mu S \partial_\mu S-\frac{1}{2}\mu^2_S \,S^2 -\frac{\lambda_S}{4} S^4 -\lambda_L\, H^\dagger H\, S^2
\end{equation} 
with $H=(h^+ \, (h+iG_0)/\sqrt{2})^T$ the Higgs doublet. The mass of $S$ is thus given by 
\begin{equation}
\label{massS}
m_S^2=\mu^2_S+\lambda_L \mbox{\rm v}^2.
\end{equation}
where $\hbox{v}=246$~GeV.
In this model the sole coupling which allows $S$   to annihilate into SM particles and to interact with nuclei
is $\lambda_L$. For the annihilation cross section, the elastic scattering cross section (normalized to one nucleon) and the ratio, we obtain
\begin{eqnarray}
\sigma(S S \rightarrow \bar{f} f) v_{rel}&=&n_c\, \frac{\lambda_L^2}{\pi}\frac{m_f^2 }{m_h^4 m_S^3}
(m_S^2-m_f^2)^{3/2}\label{sigmaSannih}\\
\sigma (S {N} \rightarrow S {N})&=&\frac{\lambda_L^2}{\pi}  \frac{\mu_r^2}{m^4_h m_{S}^2} f^2 m_{N}^2 \label{sigmaSdirectdet}\\
R\equiv\sum_f \frac{ \sigma(S S \rightarrow \bar{f} f) v_{rel}}{\sigma(S N \rightarrow  S N)}&=&\sum_f
\frac{n_c m^2_f}{f^2 m^2_N \mu^2_r}\frac{(m_S^2-m_f^2)^{3/2}}{m_S}
\label{ratioS}
\end{eqnarray}
where ${n_c}=3(1)$ for quarks (leptons), $v_{rel}=(s-4 m_S^2)^{1/2}/m_S$ is the centre of mass relative velocity between both $S$ and $\mu_r=m_S m_{N}/(m_S+m_{N})$ is the nucleon-DM reduced mass. The factor $f$ parametrizes the Higgs to nucleons coupling from the trace anomaly, $f m_{N}\equiv \langle {N}| \sum_q m_q \bar{q}q|{N}\rangle=g_{h{NN}} \mbox{\rm v}$. From the results quoted in Ref.~\cite{frange}, we take $f=0.30$ as central value,
and vary it within the rather wide range $0.14  < f <  0.66$.
As for the Yukawa couplings, $Y_i=\sqrt{2}m_i/\mbox{\rm v}$, we consider the pole masses $m_b=4.23$ GeV, $m_c=1.2$ GeV and $m_\tau=1.77$ GeV (we neglect the effects of the running of the Yukawa couplings which are expected quite moderate). 
\begin{center}
\begin{figure}
\epsfig{file=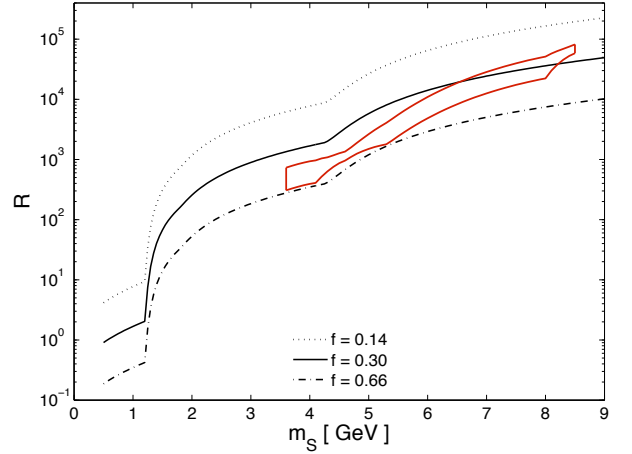,width=8.5cm}
%\vspace{-1.4mm}
\caption{Values of the ratio $R$ calculated from Eq.~(\ref{ratioS}) for three values of $f$ to be compared with the area of $R$ values required to match at $2 \sigma$ level both WMAP relic density and direct detection constraints (i.e.~DAMA, channelling effect included, and upper bounds from CDMS, Xenon, CoGent and Cresst, see Fig.~1 of Ref.~\cite{Petriello:2008jj}).}
%\vspace{-2mm}
\end{figure}
\end{center}
\vspace{-8mm}
Within the WIMP scenario, one needs $\sigma(S S \rightarrow \mbox{\rm all}\, \bar f f)  v_{rel}\sim 1 pb$ to have the right dark matter abundance. 
Therefore using Eq.(\ref{SI}), the ratio required by data is $R\simeq 10^{3}{}^{-}{}^4$. 
This has to be compared with the value of $R$ obtained from Eq.~(\ref{ratioS}). Fig.2 gives both ratios as a function of $m_S$, requiring that the relic density with respect to the critical density obtained is within the WMAP density range $ 0.094< \Omega_{DM} h^2 < 0.129$ \cite{Spergel:2006hy,Seljak:2006bg}, and that $\sigma^{SI}_p$ and $m_S$ are in the DAMA region allowed by other direct detection experiments, see Fig.~1 of Ref.~\cite{Petriello:2008jj} (taking into account the channelling effect).
As $R$ scales approximately as $m_S^2$ for $m_S \gsim m_f$, it is remarkable that both values of $R$ coincide around the DAMA DM mass region.  The relic abundances are computed using MicroMegas \cite{Belanger:2006is}\footnote{Since typically $x_f =m_{DM}/T_{fo} \approx 20$, the freeze-out temperature  is  $150$ MeV $\lsim T_{fo} \lsim 400$ MeV for DM mass in the range of (\ref{mass}). The QCD phase transition is expected to take place around $T_c \approx 200$~MeV (see for instance \cite{McLerran:2008ux}) and, consequently, for calculation of the DM abundance for DM masses as low as say $4$ GeV we should take into account the change in the expansion rate of the universe which results from the variation in number of degrees of freedom that occurs when quarks and gluons become confined in light mesons. This effect might increase the abundance by a factor of {\cal O}(2) (see for instance \cite{Bottino:2003iu}). Such an increase could be compensated by an increase of $\lambda_L$ by a factor of $\sim 1.4$ and this, in turn, would require a decrease of the parameter $f$ by the same factor. As Fig.2 shows, this may be easily accommodated  given the uncertainty on $f$. However, since MicroMegas does not take into account the QCD phase transition for the calculation of abundances, strictly speaking we are assuming here that $T_c$ is below $\sim 150$ MeV. }.  

This can also be seen in Fig.~3 which gives the regions of $m_S$ and $\lambda_L$ (for $m_h =120$~GeV) consistent with WMAP and the same direct detection constraints. Both regions nicely overlap.
The corresponding values of the SI elastic cross section can be read off from Fig.4. For the central value $f=0.30$ the overlap region covers the $m_S\approx 6$-$8$~GeV range while for $0.14<f<0.66$ regions overlap for $m_S$ between $3.5$~GeV and $8.5$~GeV.
For $f$ smaller than $0.20$ there is no overlap region.
For a fixed value of $m_S$, a smaller $f$ gives a smaller detection rate. This might be compensated by a larger  Higgs-DM coupling $\lambda_L$, but then at the expense of a smaller relic abundance.  
\begin{center}
\begin{figure}
\epsfig{file=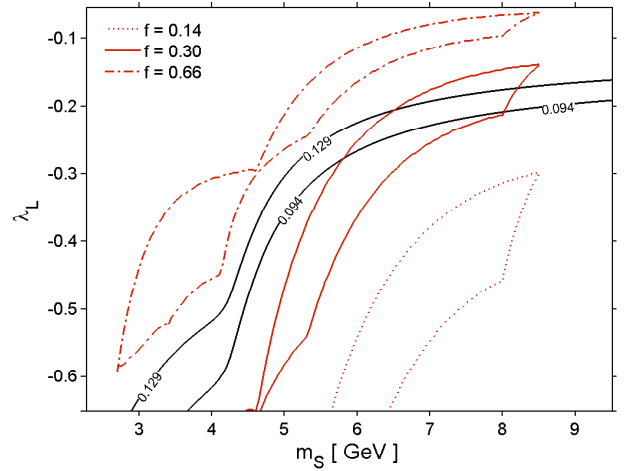,width=8.4cm}
\vspace{-2mm}
\caption{For $m_h=120$~GeV, values of $m_S$ and $\lambda_L$ which lead to the WMAP 
result, $0.094< \Omega_{DM} h^2 < 0.129$ (solid black lines), and which match the direct detection constraints ({\em i.e.}~Fig.1 of Ref.~\cite{Petriello:2008jj}), for the central value $f=0.30$ as well as the values $f=0.14$ and $f=0.66$ .}
\vspace{-2mm}
\end{figure}
\end{center}
\begin{center}
\begin{figure}
\epsfig{file=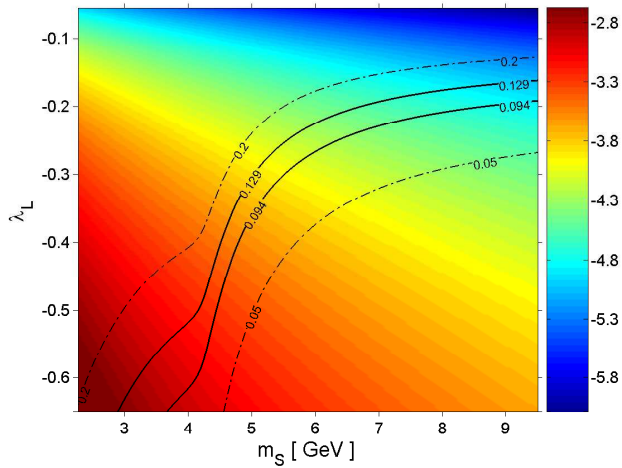,width=8.5cm}
\vspace{-2mm}
\caption{For $m_h=120$~GeV, $\log \sigma_p^{SI}$ (in pb$\equiv 10^{-36} cm^2$) as a function of $m_S$ and $\lambda_L$, versus $\Omega_{DM} h^2$ (black lines). For other values of $m_h$  $\sigma_p^{SI}$ scales as $1/m_h^4$. }
\end{figure}
\end{center}
\vspace{-15mm}
In Figs.2-4 we used $m_h=120$ GeV but
agreement  may be obtained for other Higgs boson mass provided the ratio $\lambda_L/m_h^2$ is kept fixed. 
Typically the required value is $\lambda_L/m_h^2 \simeq10^{-5}$~GeV$^{-2}$. 
To keep the result perturbative, $\lambda_L \lsim 2 \pi$,  we need that $m_h \lsim 800$~GeV. 

Since all the parameters are fixed, or strongly constrained, we may also make some definitive predictions regarding possible indirect detection of DM, in particular through gamma rays from annihilation of DM at the galactic centre.  The mass of (\ref{mass}) puts the DM candidate in the energy range of EGRET data and of the forthcoming one of GLAST. Fig.5 shows the predicted flux of gamma rays from the galactic centre for a sample of scalar DM with parameters which are consistent with DAMA and WMAP. EGRET data \cite{Hunger:1997we} are also shown \footnote{We have used the data points from around the galactic centre toward the galactic centre, left-middle plot of Fig.5c of \cite{Hunger:1997we}}.  It is interesting that the predicted flux  is of the order of magnitude of the observed flux at the lowest energies that have been probed by  EGRET. Actually, the predicted flux may even be larger than the observed one for some set of parameters. We have however refrained from putting constraints on the model parameters, given the large uncertainties on the abundance of dark matter at the centre of the Galaxy. Here we  assume a mundane NFW profile \cite{Navarro:1996gj}, as in \cite{LopezHonorez:2006gr}.
Similar predictions regarding the gamma flux,  for  a different model compatible with DAMA, have been done in \cite{Feng:2008dz}.

A number of remarks are now in order.
First of all, we should keep in mind that the parameter fit to detection data depends on precise assumptions on the abundance and velocity of the distribution of dark matter in our neighbourhood (see {\em e.g.}~\cite{Petriello:2008jj}). Also, the cosmic relic abundance is obtained assuming a mundane, radiation dominated expansion of the universe. These are conservative but well motivated options. However relaxing either one or the other would give 
other predictions and would likely jeopardize the model.

On the experimental side, it is to be seen if the intriguing results of DAMA will survive  the test of time and will be eventually confirmed by other SI experiments, 
like CRESST (see {\em i.e.}~the discussion in \cite{Petriello:2008jj}). Further confirmation of the channeling effect in the regime studied by DAMA would also be useful. Without channeling, there is no region allowed by all experiments \footnote{For an explanation of the channelling effect see \cite{Bernabei:2007hw}. For an independent analysis see \cite{Petriello:2008jj}.}\footnote{Note added in proof: For further discussions of the allowed domain of dark matter mass {\em vs} cross section, see \cite{Chang:2008xa,Fairbairn:2008gz,Savage:2008er}.}. Regarding SD detectors, as there are no (non-negligible) spin dependent (SD) interactions in the Higgs exchange scenario, no signal should be expected.  

On the theoretical side, we definitely wish we had  a better determination of the form factor $f$ in the Higgs-nucleon effective coupling
(see the discussion above). 

Also on the theoretical side, inspection of Fig.~3 and 4 reveals that substantial tuning is necessary to obtain agreement with DAMA data, as one needs  a quite
large SI cross section. As already emphasized in \cite{Burgess:2000yq} (see also \cite{Barger:2007im} for a singlet and \cite{LopezHonorez:2006gr} for the IDM\footnote{The mass range considered here is quite different from the larger ones considered in \cite{Barbieri:2006dq} and \cite{LopezHonorez:2006gr}. The main reason is that those works put more emphasis on the gauge interactions of the DM than on its coupling to the Higgs. That there are low mass solutions consistent with WMAP is however implicit in, for instance, the Fig.5 of \cite{LopezHonorez:2006gr}.})  the cross section of a scalar DM candidate is typically small, $10^{-8} pb$, below the range of all
 the existing direct detection experiments. For it to be observable in DAMA requires a large, albeit still perturbative  coupling $\vert\lambda_L\vert \sim 0.1-1$ between the DM and the Higgs \footnote{For negative couplings $\lambda_L$, stability of the potential (\ref{lag}) imposes that $\lambda_S,\lambda >0$ and $\lambda \lambda_S >\lambda_L^2$, where $\lambda$ is the Higgs quartic coupling \cite{Burgess:2000yq}, conditions which are easely to satisfy for the parameter range we consider. Similar constraints hold for the Inert Doublet Model \cite{Barbieri:2006dq}}. 
Simultaneously, to keep the DM mass within the range (\ref{mass})  requires a cancellation between both terms in Eq.~(\ref{massS}) at the  per cent level, roughly.  This, in our opinion, is not unbearable.  Nor is it suprising, given the very minimal number of parameters of the model. Taken at face value, it implies a close connection between DM and the Higgs sector, thus with the mechanism at the origin of electroweak symmetry breaking.

This brings us to one immediate, striking consequence of such DM models. A large coupling to the Higgs leads 
to a large Higgs boson decay rate to a scalar DM pair \cite{Burgess:2000yq}. For example for $m_S=7$~GeV, $\lambda_L=-0.2$ and for a Higgs of mass 120 GeV we get the branching ratio $BR(h \rightarrow SS)=99.5\,\%$, while for $m_h=200$ GeV and $\lambda_L=-0.55$ we get $BR(h \rightarrow SS)\simeq 70\,\%$. This reduces the visible branching ratio accordingly, rendering the Higgs boson basically invisible at LHC for $m_h=120$~GeV, except possibly for many years of high luminosity data taking. Such a dominance of the invisible DM channel is a clear prediction of the framework, although a challenging one for experimentalists for low value of $m_h$ (see \cite{Eboli:2000ze} for strategies to search an invisible Higgs at the LHC).
\begin{center}
\begin{figure}
\epsfig{file=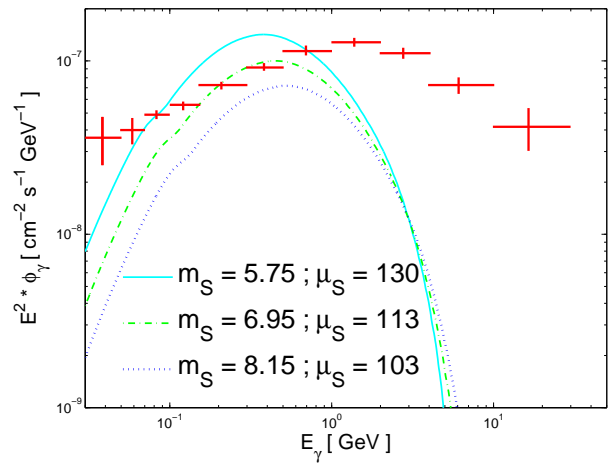,width=6.45cm,angle=270}
\caption{For three example values of $m_S$ and $\mu_2$ (in GeV) taken from Fig.~3, flux of gamma rays from the galactic centre from the annihilation of a scalar DM consistent with DAMA, compared with EGRET data.
}
\end{figure}
\end{center}
%\vspace{-8mm}
We now would like to emphasize that the results discussed here apply for any scalar dark matter model for which annihilation and SI scattering cross sections would be dominated by the diagrams of Fig.~1. A simple such an instance is the IDM, in which case the DM candidate is the lightest component of a scalar doublet, odd under a $Z_2$ symmetry. All the results above hold, provided one
identifies the appropriate parameters. Using the notations of~\cite{LopezHonorez:2006gr} for instance, we get the same abundances and SI cross section in the DAMA range, Figs.~2-4, provided one replaces $m_S \rightarrow m_{H_0}$, $\mu_S \rightarrow \mu_2$ and $\lambda_L\rightarrow \lambda_L$. One just has to make sure that the extra components of the doublets, noted $A_0$ and $H^\pm$, are heavy enough. In practice, this means that we need $m_{H_0} + m_{A_0} > m_Z$ so as not to contribute to the $Z$ invisible decay width. This simultaneoulsy kills the possibility of sizable co-annihilation of $H_0$ with $A_0$ in the early universe or elastic scattering in detectors through a $Z$ channel.  As we consider $m_{H_0}$ in the range (\ref{mass}), $m_{A_0} \gsim m_Z$. Furthermore we need to preclude large radiative corrections to the $Z$ and $W$ bosons, and in particular to the $\rho$ parameter (see the discussion in \cite{Barbieri:2006dq}). Within the IDM, this may be obtained naturally  as there is a global custodial symmetry if $m_{A_0} \approx m_{H^\pm}$,  in which case the contribution of the extra scalar doublet to  $\rho$ vanishes (see \cite{Gerard:2007kn} and the discussion in \cite{Hambye:2007vf}). Of course, the extra heavy degrees of freedom $A_0$ and $H^\pm$ might  show up eventually in high energy collisions \cite{Cao:2007rm}, something in which the IDM differs from the simpler singlet scalar model. 

For the sake of comparison with the scalar DM case discussed above, we now briefly consider the case of  singlet Dirac and Majorana fermionic DM candidates. In the former case, we consider the Lagrangian 
\begin{equation}
{\cal L} \owns  \bar{\psi}(i \partial \hspace{-0.2cm}\slash -m_{0})\psi -\frac{Y_\psi}{\sqrt{2}} \bar{\psi} \psi h
\end{equation}
The Higgs-DM coupling may arise from the non-renormalizable operator $\bar \psi \psi \,H^\dagger H/\Lambda$, so that $Y_\psi \propto \mbox{\rm v}/\Lambda$, and the mass of $\psi$ is
$m_\psi = m_{0} + Y_{\psi} \mbox{\rm v}/\sqrt{2}$ (for similar models see {\em i.e.} \cite{SungCheon:2008ts}).
For the annihilation and direct detection cross sections we get
\begin{eqnarray}
\sigma(\bar{\psi} \psi \rightarrow \bar{f} f)v_{rel}&=&  n_c\frac{Y_\psi^2}{16 \pi }\, \frac{ m_f^2 v^2_{rel}}{\mbox{\rm v}^2 m_h^4}  \frac{(m^2_\psi-m_f^2)^{3/2}}{m_\psi}  \nonumber \\
\sigma(\psi {N} \rightarrow \psi {N})&=& \frac{Y_\psi^2}{2 \pi} \frac{\mu^2_r}{\mbox{\rm v}^2 m_h^4} f^2 m_{N}^2 \label{sigmaPsidirectdet}\\
R\equiv \frac{ \sum_f \sigma(\bar{\psi} \psi \rightarrow \bar{f} f)v_{rel}}{\sigma(\psi {N} \rightarrow  \psi {N})} \label{ratioPsi} 
&=&  \frac{\sum_f n_c m^2_f}{f^2 m^2_{N} \mu^2_r}\frac{v^2_{rel}}{8}\frac{(m^2_\psi-m_f^2)^{3/2}}{m_\psi} \nonumber
\end{eqnarray}
Both the cross sections and the ratio have different parametric dependence compared to the DM scalar case, Eqs.~(\ref{sigmaSannih})-(\ref{ratioS}).  Keeping all things constant, 
the SI independent cross section is smaller by a factor of $m_{DM}^2/\mbox{\rm v}^2$. Also, assuming perturbative values of the coupling $Y_\psi$, the abundance of dark matter is way larger than the critical density [for parameters consistent with (\ref{SI}) and (\ref{mass})] because of the extra small factor   $m_{DM}^2/\mbox{\rm v}^2 \cdot v_{rel}^2$ in the annihilation cross-section compared to the scalar DM case. The origin of this suppression is well-known (see {\em e.g.} \cite{Jungman:1995df}). It results from the fact that a fermion-antifermion pair in a s-wave is a CP odd state and can not annihilate into a scalar, like the Higgs. Hence the annihilation is both p-wave, $\propto v_{rel}^2$, and helicity suppressed, $\propto m_{DM}^2$.
Similar conclusions hold for the Majorana case. In these fermion cases other channels than Higgs exchange must necessarily be present in order to match the WMAP and DAMA observations, as for instance is the case in the MSSM (see the recent discussion of \cite{Bottino:2008mf}).

Before concluding, let us raise that the scenario discussed here does not shed light on the baryon-dark matter coincidence problem, $\Omega_{DM} \sim \Omega_B$. 
If the dark matter mass is in the range (\ref{mass}), there is roughly one dark matter particle per baryon in the universe, $n_{DM} \approx n_B$, a result which   suggests a 
deeper connection between ordinary and dark matters. A few models have been proposed to naturally explain this coincidence (see {\em e.g.}
\cite{Kaplan:1991ah,Barr:1991qn,Farrar:2005zd,Kitano:2004sv,Cosme:2005sb,Abel:2006hr,McDonald:2006if,Hooper:2004dc,Thomas:1995ze,Dodelson:1989cq}). 
They tend to be sophisticated in comparison with the model discussed here and, to our knowledge, 
they also tend to have cross sections for scattering of dark matter with nuclei that are much below  the sensitivity of existing and forthcoming detectors.

\bigskip
In conclusion, we have discussed the possible relevance of Higgs exchange for WIMP in  DAMA mass and SI cross section ranges, leading to a one-to-one relation between the relic density and SI direct detection rate. 
We have emphasized that for scalar dark matter this relation is in agreement with data, provided Higgs exchange is the dominant channel in both annihilation and SI cross sections.
We have
illustrated this through a singlet real scalar and the inert doublet model, where it is naturally realized. These models are very simple, give definitive predictions --including for the LHC-- and might be falsifiable within a foreseeable future.

\vspace{5mm}
\section*{Acknowledgement}
\vspace{-3mm}
 We thank Jean-Marie Fr\`ere for stimulating discussions. One of us (S.A.)  thanks  Michael Kr\"amer. This work is supported by the FNRS, the IISN and the Belgian Federal Science Policy (IAP VI/11). Preprint ULB-TH/08-27.


\begin{thebibliography}{99}


\bibitem{DAMA}
DAMA collaboration, R.~Bernabei et al., arXiv:0804.2741 [astro-ph].

\bibitem{DAMA03}
DAMA collaboration, R.~Bernabei et al., Riv. Nuovo Cim. {\bf 26N1} (2003) 1; Int. J. Mod. Phys. {\bf D13} (2004) 2127.

%\cite{Akerib:2003px}
\bibitem{CDMSXenon}
  D.~S.~Akerib {\it et al.}  [CDMS Collaboration],
  %``New results from the Cryogenic Dark Matter Search experiment,''
  Phys.\ Rev.\  D {\bf 68}, 082002 (2003)
  [arXiv:hep-ex/0306001];
  %%CITATION = PHRVA,D68,082002;%%
    %\cite{Ahmed:2008eu}
%\bibitem{CDMSII}
  Z.~Ahmed {\it et al.}  [CDMS Collaboration],
  %``A Search for WIMPs with the First Five-Tower Data from CDMS,''
  arXiv:0802.3530 [astro-ph];
  %%CITATION = ARXIV:0802.3530;%%
     %\cite{Akerib:2005kh}
%\bibitem{CDMSIISi}
  D.~S.~Akerib {\it et al.}  [CDMS Collaboration],
  %``Limits on spin-independent WIMP nucleon interactions from the two-tower
  %run of the Cryogenic Dark Matter Search,''
  Phys.\ Rev.\ Lett.\  {\bf 96}, 011302 (2006)
  [arXiv:astro-ph/0509259];
  %%CITATION = PRLTA,96,011302;%%
   %\cite{Angle:2007uj}
%\bibitem{XENON}
  J.~Angle {\it et al.}  [XENON Collaboration],
  %``First Results from the XENON10 Dark Matter Experiment at the Gran Sasso
 %National Laboratory,''
  Phys.\ Rev.\ Lett.\  {\bf 100}, 021303 (2008)
  [arXiv:0706.0039 [astro-ph]].
  %%CITATION = PRLTA,100,021303;%%


%\cite{Petriello:2008jj}
\bibitem{Petriello:2008jj}
  F.~Petriello and K.~M.~Zurek,
  %``DAMA and WIMP dark matter,''
  arXiv:0806.3989 [hep-ph].
  %%CITATION = ARXIV:0806.3989;%%

%\cite{Foot:2008nw}
\bibitem{Foot:2008nw}
  R.~Foot,
  %``Mirror dark matter and the new DAMA/LIBRA results: A simple explanation for
  %a beautiful experiment,''
  arXiv:0804.4518 [hep-ph].
  %%CITATION = ARXIV:0804.4518;%%

%\cite{Feng:2008dz}
\bibitem{Feng:2008dz}
  J.~L.~Feng, J.~Kumar and L.~E.~Strigari,
  %``Explaining the DAMA Signal with WIMPless Dark Matter,''
  arXiv:0806.3746 [hep-ph].
  %%CITATION = ARXIV:0806.3746;%%

%\cite{Bottino:2008mf}
\bibitem{Bottino:2008mf}
  A.~Bottino, F.~Donato, N.~Fornengo and S.~Scopel,
  %``Interpreting the recent results on direct search for dark matter particles
  %in terms of relic neutralino,''
  arXiv:0806.4099 [hep-ph].
  %%CITATION = ARXIV:0806.4099;%%

%\cite{Avignone:2008cw}
\bibitem{Avignone:2008cw}
  F.~T.~Avignone, R.~J.~Creswick and S.~Nussinov,
  %``Searching direction dependent daily modulation in dark matter detectors,''
  arXiv:0807.3758 [hep-ph].
  %%CITATION = ARXIV:0807.3758;%%


%\cite{Yao:2006px}
\bibitem{Yao:2006px}
  W.~M.~Yao {\it et al.}  [Particle Data Group],
  %``Review of particle physics,''
  J.\ Phys.\ G {\bf 33} (2006) 1.
  %%CITATION = JPHGB,G33,1;%% 


%\cite{Deshpande:1977rw}
\bibitem{Deshpande:1977rw}
  N.~G.~Deshpande and E.~Ma,
  %`Pattern Of Symmetry Breaking With Two Higgs Doublets,''
  Phys.\ Rev.\ D {\bf 18} (1978) 2574.
  %%CITATION = PHRVA,D18,2574;%%



 

  %\cite{Ma:2006km} 
\bibitem{Ma:2006km} 
  E.~Ma, 
  %``Verifiable radiative seesaw mechanism of neutrino mass and dark matter,'' 
  Phys.\ Rev.\ D {\bf 73} (2006) 077301. 
%  [arXiv:hep-ph/0601225]. 
  %%CITATION = HEP-PH 0601225;%% 


 
%\cite{Barbieri:2006dq} 
\bibitem{Barbieri:2006dq} 
  R.~Barbieri, L.~J.~Hall and V.~S.~Rychkov, 
   %``Improved naturalness with a heavy Higgs: An alternative road to LHC 
  %physics,'' 
  Phys.\ Rev.\ D {\bf 74} (2006) 015007 
  [arXiv:hep-ph/0603188]. 
  %%CITATION = HEP-PH 0603188;%% 




  %\cite{Majumdar:2006nt} 
\bibitem{Majumdar:2006nt} 
  D.~Majumdar and A.~Ghosal, 
  %``Dark matter candidate in inert doublet model: Direct detection rates,'' 
  arXiv:hep-ph/0607067. 
  %%CITATION = HEP-PH 0607067;%% 
  %\cite{Kubo:2006yx} 
 

%\cite{LopezHonorez:2006gr}
\bibitem{LopezHonorez:2006gr}
  L.~Lopez Honorez, E.~Nezri, J.~F.~Oliver and M.~H.~G.~Tytgat,
  %``The inert doublet model: An archetype for dark matter,''
  JCAP {\bf 0702} (2007) 028
 % [arXiv:hep-ph/0612275].
  %%CITATION = JCAPA,0702,028;%%

%\cite{Gustafsson:2007pc}
\bibitem{Gustafsson:2007pc}
  M.~Gustafsson, E.~Lundstrom, L.~Bergstrom and J.~Edsjo,
  %``Significant gamma lines from inert Higgs dark matter,''
  arXiv:astro-ph/0703512.
  %%CITATION = ASTRO-PH/0703512;%%


%\cite{Hambye:2007vf}
\bibitem{Hambye:2007vf}
  T.~Hambye and M.~H.~G.~Tytgat,
  %``Electroweak Symmetry Breaking induced by Dark Matter,''
  Phys.\ Lett.\  B {\bf 659} (2008) 651
  [arXiv:0707.0633 [hep-ph]].
  %%CITATION = PHLTA,B659,651;%%

\bibitem{McDonald1} 
J.~McDonald, Phys.\ Rev.\  {\bf D50} (1994) 3637.

\bibitem{McDonald:2001vt}
  J.~McDonald,
  %``Thermally generated gauge singlet scalars as self-interacting dark
 % %matter,''
  Phys.\ Rev.\ Lett.\  {\bf 88} (2002) 091304.
%arXiv:hep-ph/0106249.
  %%%CITATION = PRLTA,88,091304;%%
%%  \bibitem{Boehm} C. Boehm, , and refs. therein.

%\cite{Burgess:2000yq}
\bibitem{Burgess:2000yq}
  C.~P.~Burgess, M.~Pospelov and T.~ter Veldhuis,
  %``The minimal model of nonbaryonic dark matter: A singlet scalar,''
  Nucl.\ Phys.\  B {\bf 619} (2001) 709
  [arXiv:hep-ph/0011335].
  %%CITATION = NUPHA,B619,709;%%

%\cite{Patt:2006fw}
\bibitem{Patt:2006fw}
  B.~Patt and F.~Wilczek,
  %``Higgs-field portal into hidden sectors,''
  arXiv:hep-ph/0605188.
  %%CITATION = HEP-PH/0605188;%%

%\cite{Meissner:2006zh}
\bibitem{Meissner:2006zh}
  K.~A.~Meissner and H.~Nicolai,
  %``Conformal symmetry and the standard model,''
  Phys.\ Lett.\  B {\bf 648} (2007) 312
  [arXiv:hep-th/0612165].
  %%CITATION = PHLTA,B648,312;%%

%\cite{Barger:2007im}
\bibitem{Barger:2007im}
  V.~Barger, P.~Langacker, M.~McCaskey, M.~J.~Ramsey-Musolf and G.~Shaughnessy,
  %``LHC Phenomenology of an Extended Standard Model with a Real Scalar
  %Singlet,''
  Phys.\ Rev.\  D {\bf 77} (2008) 035005
  [arXiv:0706.4311 [hep-ph]].
  %%CITATION = PHRVA,D77,035005;%%

%\cite{Espinosa:2007qk}
\bibitem{Espinosa:2007qk}
  J.~R.~Espinosa and M.~Quiros,
  %``Novel effects in electroweak breaking from a hidden sector,''
  Phys.\ Rev.\  D {\bf 76} (2007) 076004
  [arXiv:hep-ph/0701145].
  %%CITATION = PHRVA,D76,076004;%%

%\cite{SungCheon:2007nw}
\bibitem{SungCheon:2007nw}
  H.~Sung Cheon, S.~K.~Kang and C.~S.~Kim,
  %``Low Scale Leptogenesis and Dark Matter Candidates in an Extended Seesaw
  %Model,''
  JCAP {\bf 0805} (2008) 004
  [arXiv:0710.2416 [hep-ph]].
  %%CITATION = JCAPA,0805,004;%%


%\cite{Bottino:2003iu}
\bibitem{Bottino:2003iu}
  A.~Bottino, F.~Donato, N.~Fornengo and S.~Scopel,
  %``Lower bound on the neutralino mass from new data on CMB and  implications
  %for relic neutralinos,''
  Phys.\ Rev.\  D {\bf 68} (2003) 043506
  [arXiv:hep-ph/0304080].
  %%CITATION = PHRVA,D68,043506;%%


\bibitem{frange}
M.M. Pavan, R.A. Arndt, I.I. Strakovski and R.L. Workman, PiN Newslett. {\bf 16} (2002) 110; A. Bottino, F. Donato, N. Fornengo and S. Scopel, Astroparticle Physics {\bf 13} (2000) 215; R.~Koch, Z. Phys. {\bf C15} (1982) 161; J.~Gasser, H.~Leutwyler and M.E. Sainio, Phys. Lett. {\bf B253} (1991) 260.
  


%\cite{Spergel:2006hy}
\bibitem{Spergel:2006hy}
  D.~N.~Spergel {\it et al.},
  %``Wilkinson Microwave Anisotropy Probe (WMAP) three year results:
  %Implications for cosmology,''
  arXiv:astro-ph/0603449.
  %%CITATION = ASTRO-PH 0603449;%% 
%\cite{Seljak:2006bg}
\bibitem{Seljak:2006bg}
  U.~Seljak, A.~Slosar and P.~McDonald,
  %``Cosmological parameters from combining the Lyman-alpha forest with CMB,
  %galaxy clustering and SN constraints,''
  JCAP {\bf 0610} (2006) 014.
%  [arXiv:astro-ph/0604335].
  %%CITATION = ASTRO-PH 0604335;%%


%\cite{Belanger:2006is}
\bibitem{Belanger:2006is}
  G.~Belanger, F.~Boudjema, A.~Pukhov and A.~Semenov,
  %``micrOMEGAs2.0: A program to calculate the relic density of dark matter  in
  %a generic model,''
  Comput.\ Phys.\ Commun.\  {\bf 176} (2007) 367.
  %[arXiv:hep-ph/0607059].
  %%CITATION = CPHCB,176,367;%%

%\cite{Bernabei:2007hw}
\bibitem{Bernabei:2007hw}
  R.~Bernabei {\it et al.},
  %``Possible implications of the channeling effect in NaI(Tl) crystals,''
  Eur.\ Phys.\ J.\  C {\bf 53} (2008) 205
  [arXiv:0710.0288 [astro-ph]].
  %%CITATION = EPHJA,C53,205;%%



%\cite{Gerard:2007kn}
\bibitem{Gerard:2007kn}
  J.~M.~G\'erard and M.~Herquet,
  %``A twisted custodial symmetry in the two-Higgs-doublet model,''
  Phys.\ Rev.\ Lett.\  {\bf 98} (2007) 251802
  [arXiv:hep-ph/0703051].
  %%CITATION = PRLTA,98,251802;%%

%\cite{Cao:2007rm}
\bibitem{Cao:2007rm}
  Q.~H.~Cao, E.~Ma and G.~Rajasekaran,
  %``Observing the Dark Scalar Doublet and its Impact on the Standard-Model
  %Higgs Boson at Colliders,''
  Phys.\ Rev.\  D {\bf 76} (2007) 095011
  [arXiv:0708.2939 [hep-ph]].
  %%CITATION = PHRVA,D76,095011;%%

%\cite{SungCheon:2008ts}
\bibitem{SungCheon:2008ts}
  H.~Sung Cheon, S.~K.~Kang and C.~S.~Kim,
  %``Doubly Coexisting Dark Matter Candidates in an Extended Seesaw Model,''
  arXiv:0807.0981 [hep-ph].
  %%CITATION = ARXIV:0807.0981;%%

%\cite{Jungman:1995df}
\bibitem{Jungman:1995df}
  G.~Jungman, M.~Kamionkowski and K.~Griest,
  %``Supersymmetric dark matter,''
  Phys.\ Rept.\  {\bf 267} (1996) 195
  [arXiv:hep-ph/9506380].
  %%CITATION = PRPLC,267,195;%%

%\cite{Kaplan:1991ah}
\bibitem{Kaplan:1991ah}
  D.~B.~Kaplan,
  %``A Single explanation for both the baryon and dark matter densities,''
  Phys.\ Rev.\ Lett.\  {\bf 68} (1992) 741.
  %%CITATION = PRLTA,68,741;%%

%\cite{Barr:1991qn}
\bibitem{Barr:1991qn}
  S.~M.~Barr,
  %``Baryogenesis, sphalerons and the cogeneration of dark matter,''
  Phys.\ Rev.\  D {\bf 44} (1991) 3062.
  %%CITATION = PHRVA,D44,3062;%%


%\cite{Farrar:2005zd}
\bibitem{Farrar:2005zd}
  G.~R.~Farrar and G.~Zaharijas,
  %``Dark matter and the baryon asymmetry,''
  Phys.\ Rev.\ Lett.\  {\bf 96} (2006) 041302
  [arXiv:hep-ph/0510079].
  %%CITATION = PRLTA,96,041302;%%

%\cite{Kitano:2004sv}
\bibitem{Kitano:2004sv}
  R.~Kitano and I.~Low,
  %``Dark matter from baryon asymmetry,''
  Phys.\ Rev.\  D {\bf 71} (2005) 023510
  [arXiv:hep-ph/0411133].
  %%CITATION = PHRVA,D71,023510;%%

%\cite{Cosme:2005sb}
\bibitem{Cosme:2005sb}
  N.~Cosme, L.~Lopez Honorez and M.~H.~G.~Tytgat,
  %``Leptogenesis and dark matter related?,''
  Phys.\ Rev.\  D {\bf 72} (2005) 043505
  [arXiv:hep-ph/0506320].
  %%CITATION = PHRVA,D72,043505;%%

%\cite{Abel:2006hr}
\bibitem{Abel:2006hr}
  S.~Abel and V.~Page,
  %``Affleck-Dine (pseudo)-Dirac neutrinogenesis,''
  JHEP {\bf 0605} (2006) 024
  [arXiv:hep-ph/0601149].
  %%CITATION = JHEPA,0605,024;%%

%\cite{McDonald:2006if}
\bibitem{McDonald:2006if}
  J.~McDonald,
  %``Right-handed sneutrino condensate cold dark matter and the  baryon-to-dark
  %matter ratio,''
  JCAP {\bf 0701} (2007) 001
  [arXiv:hep-ph/0609126].
  %%CITATION = JCAPA,0701,001;%%

%\cite{Hooper:2004dc}
\bibitem{Hooper:2004dc}
  D.~Hooper, J.~March-Russell and S.~M.~West,
  %``Asymmetric sneutrino dark matter and the Omega(b)/Omega(DM) puzzle,''
  Phys.\ Lett.\  B {\bf 605} (2005) 228
  [arXiv:hep-ph/0410114].
  %%CITATION = PHLTA,B605,228;%%

%\cite{Thomas:1995ze}
\bibitem{Thomas:1995ze}
  S.~D.~Thomas,
  %``Baryons and dark matter from the late decay of a supersymmetric
  %condensate,''
  Phys.\ Lett.\  B {\bf 356} (1995) 256
  [arXiv:hep-ph/9506274].
  %%CITATION = PHLTA,B356,256;%%

%\cite{Dodelson:1989cq}
\bibitem{Dodelson:1989cq}
  S.~Dodelson and L.~M.~Widrow,
  %``BARYOGENESIS IN A BARYON SYMMETRIC UNIVERSE,''
  Phys.\ Rev.\  D {\bf 42} (1990) 326.
  %%CITATION = PHRVA,D42,326;%%

%\cite{Navarro:1996gj}
\bibitem{Navarro:1996gj}
  J.~F.~Navarro, C.~S.~Frenk and S.~D.~M.~White,
  %``A Universal Density Profile from Hierarchical Clustering,''
  Astrophys.\ J.\  {\bf 490} (1997) 493
  [arXiv:astro-ph/9611107].
  %%CITATION = ASJOA,490,493;%%

%\cite{McLerran:2008ux}
\bibitem{McLerran:2008ux}
  L.~McLerran,
  %``Quarkyonic Matter and the Phase Diagram of QCD,''
  arXiv:0808.1057 [hep-ph].
  %%CITATION = ARXIV:0808.1057;%%

%\cite{Hunger:1997we}
\bibitem{Hunger:1997we}
  S.~D.~Hunger {\it et al.},
  %``EGRET observations of the diffuse gamma-ray emission from the galactic
  %plane,''
  Astrophys.\ J.\  {\bf 481} (1997) 205.
  %%CITATION = ASJOA,481,205;%%

%\cite{Bergstrom:1997fj}
\bibitem{Bergstrom:1997fj}
  L.~Bergstrom, P.~Ullio and J.~H.~Buckley,
  %``Observability of gamma rays from dark matter neutralino annihilations  in
  %the Milky Way halo,''
  Astropart.\ Phys.\  {\bf 9} (1998) 137
  [arXiv:astro-ph/9712318].
  %%CITATION = APHYE,9,137;%%
 

%\cite{Eboli:2000ze}
\bibitem{Eboli:2000ze}
  O.~J.~P.~Eboli and D.~Zeppenfeld,
  %``Observing an invisible Higgs boson,''
  Phys.\ Lett.\  B {\bf 495} (2000) 147
  [arXiv:hep-ph/0009158].
  %%CITATION = PHLTA,B495,147;%%

%\cite{Chang:2008xa}
\bibitem{Chang:2008xa}
  S.~Chang, A.~Pierce and N.~Weiner,
  %``Using the Energy Spectrum at DAMA/LIBRA to Probe Light Dark Matter,''
  arXiv:0808.0196 [hep-ph].
  %%CITATION = ARXIV:0808.0196;%%


%\cite{Fairbairn:2008gz}
\bibitem{Fairbairn:2008gz}
  M.~Fairbairn and T.~Schwetz,
  %``Spin-independent elastic WIMP scattering and the DAMA annual modulation
  %signal,''
  arXiv:0808.0704 [hep-ph].
  %%CITATION = ARXIV:0808.0704;%%

%\cite{Savage:2008er}
\bibitem{Savage:2008er}
  C.~Savage, G.~Gelmini, P.~Gondolo and K.~Freese,
  %``Compatibility of DAMA/LIBRA dark matter detection with other searches,''
  arXiv:0808.3607 [astro-ph].
  %%CITATION = ARXIV:0808.3607;%%



\end{thebibliography}
\end{document}